\documentclass[english,
]{aa}
\usepackage{epsf,amsfonts,amssymb,graphicx,fancyheadings,caption}
\usepackage{natbib}
\usepackage{babel}
\usepackage{txfonts}
\usepackage{rotating}
\usepackage{enumitem}

\bibpunct{(}{)}{;}{a}{}{,}   

\begin{document}

\title{HdC and EHe stars through the prism of Gaia DR3: }
\subtitle{Evolution of RV amplitude and dust formation rate with effective temperature}

\author{
P.~Tisserand\inst{1},
C.~L.~Crawford\inst{2},
J.~Soon\inst{3},
G.~C.~Clayton\inst{4},
A.~J.~Ruiter\inst{5},
I.~R.~Seitenzahl\inst{5}
}

\institute{
Sorbonne Universit\'es, UPMC Univ Paris 6 et CNRS, UMR 7095, Institut d'Astrophysique de Paris, IAP, F-75014 Paris, France \and
Sydney Institute for Astronomy (SIfA), School of Physics, University of Sydney, NSW 2006, Australia \and
Research School of Astronomy and Astrophysics, Australian National University, Cotter Rd, Weston Creek ACT 2611, Australia \and
Department of Physics \& Astronomy, Louisiana State University, Baton Rouge, LA 70803, USA \and
School of Science, University of New South Wales, Canberra, ACT 2600, Australia
}


\offprints{Patrick Tisserand; \email{tisserand@iap.fr}}

\date{}


\abstract {The Gaia DR3 release includes heliocentric radial velocity measurements and velocity variability indices for tens of millions of stars observed over 34 months.}
{In this study, we utilise these indices to investigate the intrinsic radial velocity variations of Hydrogen-deficient Carbon (HdC) stars and Extreme Helium (EHe) stars across their large ranges of temperature and brightness.}
{Taking advantage of the newly defined HdC temperature classes, we examine the evolution of the total velocity amplitude with effective temperature. Additionally, we analyse the variation in the dust production rate of R Coronae Borealis (RCB) stars with temperature using two different proxies for the photometric state of RCB stars: one from Gaia and another from the 2MASS survey.}
{Our observations revealed a trend in the evolution of the maximum radial velocity amplitude across each HdC temperature class. Similarly, we also observed a correlation between stellar temperature and the dust production rate.}
{Interestingly, we possibly observed for the first time some variations of the intrinsic radial velocity amplitude and the dust production rate with HdC temperature class. If confirmed, these variations would indicate that the helium shell-burning giant stage starts with strong atmospheric motions that decrease in strength, up to $\sim$6000 K, before picking up again as the HdC star atmosphere shrinks further in size and reaches warmer temperatures. Moreover, the dust formation rate appears to be much higher in colder RCB stars compared to warmer ones.}


\authorrunning{Tisserand, P. }
\titlerunning{Evolution of RV amplitude and dust formation rate with effective temperature}

\maketitle

\section{Introduction \label{sec_intro}}

Hydrogen-deficient Carbon (HdC) stars are rare supergiant stars with an effective temperature (T$_{eff}$) ranging between 3500 and 8500 K \citep{2023MNRAS.521.1674C}. Accumulating evidence suggests a cataclysmic origin for these stars, which was originally theorised by \citet{1984ApJ...277..355W,1998MNRAS.296.1019H}. Indeed, they are strongly suspected to result from the merger of one CO- and one He-rich white dwarf \citep{2012JAVSO..40..539C, 2011MNRAS.414.3599J} with a combined total mass between 0.6 and 1.1 solar masses \citep[see Fig.13]{2022A&A...667A..83T}, as predicted by population synthesis simulations of close binary systems \citep{2015ApJ...809..184K} after various mass transfer phases. 

The HdC stars are categorised into two groups based on their dust production: the R Coronae Borealis (RCB) stars, known for their rapid and unpredictable photometric declines (up to 9 mag in V) caused by the emergence of newly formed dust clouds on the line of sight, and the dustless HdC (dLHdC) stars. However, distinguishing between RCB and dLHdC stars is not always straightforward. Some dLHdC stars, including F75, F152, C526, and A166, have been observed to produce a minor amount of dust, indicated by a weak IR excess signal \citep{2022A&A...667A..83T}. Hence, it is conceivable that certain dLHdC stars could be RCB stars that have temporarily halted dust production.

Typically, RCB stars exhibit an absolute magnitude within the range of -5 $<M_V<$ -3 mag, whereas dLHdC stars have been observed to be less luminous, on average, by approximately 2 mag \citep{2022A&A...667A..83T}. This suggests dLHdC stars may have originated from white dwarf mergers of lower total mass. Spectroscopically, both groups share very similar characteristics, even if \citet{2022A&A...667A..83T} observed notable differences in nitrogen and hydrogen abundances in their low-resolution spectra. Indeed, in dLHdC stars' atmosphere, these abundances appear to be respectively lower and higher than in RCB stars' atmosphere, necessitating further confirmation through high-resolution spectroscopic follow-up observations. Additionally, dLHdC stars may possess an even lower O$^{16}$/O$^{18}$ isotopic ratio \citep{2022A&A...667A..84K} compared to RCB stars \citep{2007ApJ...662.1220C, 2010ApJ...714..144G}, which is already near unity, approximately 500 times lower than the oxygen isotopic ratio measured in the sun.

Extreme Helium (EHe) stars represent another category of supergiant stars characterised by atmospheres that are nearly devoid of hydrogen. They are hotter (T$_{eff}>$8500 K) than HdC stars, suggesting they are in an evolutionary phase following the helium shell-burning giant stage (i.e., HdC stars) before transitioning into heavy white dwarfs \citep{2002MNRAS.333..121S, 2014MNRAS.445..660Z, 2019ApJ...885...27S}. On the cooler side (T$_{eff}<$3500 K), there is a growing suspicion that DYPer type stars, with DY Persei as the prototype, are associated with HdC stars \citep{2007A&A...472..247Z,2018ApJ...854..140B,2023ApJ...948...15G}. However, detailed spectroscopic studies of DY Persei in the visible and infrared have revealed that it is less hydrogen deficient than HdC stars \citep{2009ARep...53..187Y,2013ApJ...773..107G}. The DYPer type stars also produce dust but experience slower and shallower photometric declines compared to RCB stars.

At maximum brightness, RCB stars demonstrate notable photometric variations caused by pulsations, with an amplitude spanning from 0.2 to 0.4 mag and periods ranging between 30 and 100 days \citep{1994MNRAS.271..919L,1997MNRAS.285..266L,2001ApJ...554..298A}. Conversely, dLHdC stars exhibit smaller photometric variations, typically around 0.1 mag, when they are detectable \citep{2022A&A...667A..83T}. EHe stars, on the other hand, display irregular variations on shorter timescales, varying from 0.1 to approximately 25 days, with amplitudes generally less than 0.1 mag \citep{2020MNRAS.495L.135J}. Lastly, DYPer type stars show irregular photometric variations, with a total amplitude ranging between 0.1 and 0.4 mag and a typical timescale of 20 to 100 days \citep{2009A&A...501..985T}.

The first reported long-term extensive monitoring of an RCB star both photometrically and spectroscopically was published by \citet{1972MNRAS.158..305A}. They analysed the variations of RY Sgr  between 1967 and 1970 during a deep decline and a rising phase. The radial velocity (RV) measurements accumulated during the rising phase showed some large variations, with a total amplitude of about 25 km s$^{-1}$, confirming that the observed large periodic photometric oscillations of 0.5 mag amplitude and of $\sim$39 days are due to radial pulsation. This result was later confirmed by \citet{1993MNRAS.265..351L,1997MNRAS.285..266L}, who studied the photometric and RV variations of 18 HdC stars and six cool EHe stars. They showed that large pulsations such as those in RY Sgr are rare and that there exists a wide range of behaviour. Some stars, such as RT Nor, even present a similar photometric amplitude to RY Sgr, but with a much lower RV amplitude. In summary, of the 14 RCB stars studied, ten presented an RV amplitude between 10 and 20 km s$^{-1}$, while contrarily, no significant variability was found in dLHdC stars ($<$5 km s$^{-1}$). \citet{1997MNRAS.285..266L} concluded that the absence of dust formation in dLHdC stars may be due to their lower pulsation amplitude. The cool EHe stars have shown an RV amplitude similar to RCB stars (up to 20 km s$^{-1}$), despite having much lower photometric variations in V. \citet{1987MNRAS.226..317J} confirmed the scale of the RV amplitude with two other EHe stars and linked these variations to the pulsations observed in these stars.

The long-term RV variations of R CrB were studied by \citet{2019MNRAS.482.4174F} using four datasets covering the period between 1950 and 2007. The authors found that although R CrB does often exhibit light and velocity variations with a characteristic timescale of about 40 days, no coherent periodicity was detected in any of the datasets studied. The amplitudes and shapes of the velocity curves can vary significantly from cycle to cycle, which suggests that a coherent pulsation model is not applicable to R CrB.

Regarding dust formation in RCB stars, it is currently believed that dust is created in association with large turbulent flows resulting from convective cells or the entire atmosphere in the case of RY Sgr \citep{2019MNRAS.482.4174F}. Shock waves create the very localised conditions near the stellar surface (1.5-3 R$_\star$) for a sufficient amount of time to allow for the formation of dust particles, which are then expelled by radiation pressure to form dust clouds \citep{1996A&A...313..217W,1996PASP..108..225C}. Evidence of the existence of such shock waves has been exposed once by \citet{1994ApJ...432..785C} following their spectroscopic monitoring of RY Sgr throughout a pulsation cycle. If these clouds form along our line of sight, they can obscure the photosphere. Ultimately, all dust clouds merge with the surrounding circumstellar dust shell at velocities between 200 and 400 km s$^{-1}$ \citep{2011ApJ...743...44C,2011ApJ...739...37G,2018AJ....156..148M}. It is worth noting that the exact details of the dust formation process in HdC stars are still not fully understood, and ongoing research in this area aims to gain a deeper understanding of the mechanisms involved.

With DR3, the Gaia survey has released RV measurements and some RV variability indices for over 33 million sources down to GRVSmag\footnote{A narrow-band Vega-system magnitude defined from the transmittance of the Radial Velocity Spectrometer (RVS) \citep{2023A&A...674A...6S}} $\sim$14 magnitude \citep{2021A&A...653A.160S,2023A&A...674A...5K}, covering a temperature range between 3100 and 14500 K for stars brighter than the 12th magnitude and between 3100 and 6750 K for fainter stars \citep{2023A&A...674A...5K}. Multiple epochs were observed, and a combined RV was calculated as the median of the time series for the brighter stars and as an average of cross-correlation functions for the faintest stars (see their Eq.2). The resulting RV precision achieved for these two samples of stars is approximately 1.3 and 6.4 km s$^{-1}$. The RV time series have not been published yet. 

This article is published in association with another study that focuses on the distances, kinematics, and Galactic distribution of HdC, EHe, and DYPer type stars \citep{Tiss2023b}. Here, the main objective is to analyse the intrinsic velocity variability of these stars (discussion in Sect.~\ref{sec_rv}). In Section~\ref{sec_RVamp_dust_temperature}, we present our observations on the evolution of the total RV amplitude and the dust production rate with the temperature class of HdC stars. Finally, we summarise and discuss our results in Section~\ref{sec_concl}.

\section{Radial velocities \label{sec_rv}}

We analysed the published Gaia RVS parameters (see Table B.1 in \citet{2023A&A...674A...5K}) in order to investigate possible biases arising from the wide range of effective temperatures as well as from their characteristic changes in brightness and colour. Velocity variability indices were used to examine the intrinsic RV variations in the atmospheres of HdC, EHe, and DYPer type stars.

\subsection{Observed targets}



We studied the results published in the Gaia DR3 release for the following targets of interest: 

\begin{description}[font=-]
\item 129 and 34 known Galactic RCB and dLHdC stars listed in \citet[Tab.3]{2023MNRAS.521.1674C}. We added the only three known hot RCB stars, DY Cen, MV Sgr, and V348 Sgr, to the known Galactic RCB stars list. 
\item 16 Galactic EHe stars listed in \citet[Tab.1]{2011MNRAS.414.3599J} and the two EHe stars, A208 and A798, revealed in \citet{2020A&A...635A..14T}.
\item Four Galactic DYPer type stars, DY Persei itself, ASAS-DYPER-1 and -2 \citep{2013A&A...551A..77T}, and EROS2-CG-RCB-2 \citep{2008A&A...481..673T}, which we named 2MASS J17524872-2845190 (abbreviated 2MASS J175248 hereafter) to avoid any confusion with the RCB group of stars.
\item 28 known Magellanic RCB stars (see the complete list in \citet{2023MNRAS.521.1674C}). To that list, we added two known hot RCB stars, MACHO 11.8632.2507 \citep{1996ApJ...470..583A} and WISE J053745.70-635330.8 \citep{2020A&A...635A..14T}. 
\item 23 Magellanic DYPer type stars \citep{2004A&A...424..245T,2009A&A...501..985T}.
\end{description}

\subsection{Intrinsic radio velocity variabilities observed in Gaia DR3 \label{subsec_IntrinsicRVvar}}

The Gaia DR3 release includes RV variability indices for stars brighter than the 12th magnitude in GRVS. One such parameter, the P-value (rv\_chisq\_pvalue), provides an indication of the stability of the radial velocity time series \citep{2023A&A...674A...5K}. It ranges from zero to one, where lower values correspond to a higher probability of intrinsic RV variability. Stars with P-values close to one have velocity scatter consistent with measurement errors. Of the 69 bright Galactic targets published with RV measurements, 61 have an rv\_chisq\_pvalue below 0.5, indicating that they most likely have an RV time series with intrinsic variabilities (54 of which even have a value lower than 0.1). Among the other eight targets that are thus considered as possibly stable, four stand out due to their high P-value ($>$0.85), their number of visibility periods (i.e., the number of groups of observations separated by at least four days), and their number of transits, which are respectively greater than nine and ten. These targets are the dLHdC stars A226 and A249, the RCB star WISE J194218.38-203247.5, and the EHe star V4732 Sgr. No RV variability at the resolution of Gaia DR3 was detected over a total time span between $\sim$800 to 1000 days. The resulting errors on the median heliocentric RV values for these four stars range from 0.3 to 1.3 km s$^{-1}$. We note that WISE J194218.38-203247.5 was actually going through a phase of high activity in dust production, as deduced from the numerous declines observed in its ASAS-SN \citep{2014AAS...22323603S} and Catalina \citep{2015A&A...575A...2L} light curves. A decline around JD$\sim$2457000 is also detectable in its Gaia light curve.


Another interesting parameter is the peak-to-peak RV amplitude, denoted as rv\_amplitude\_robust and referred to as RVamp hereafter. It is measured over the entire RV time series after removing spurious observations. This parameter allowed us to easily compare the extent of variability for each star and to study its correlation with other physical parameters. At first glance, we noticed that all stars, except for one, had an RVamp value lower than 30 km s$^{-1}$.

The exception was the dLHdC star F152, which was reported as having an extraordinarily high RVamp value of 471 km s$^{-1}$. It was calculated from 57 Gaia transits over 24 visibility periods, which make F152 the most visited star among all our targets. F152 is located in the Galactic halo, towards the edge of the Large Magellanic Cloud, and it has an effective temperature ranging between 6800-7800 K \citep{2023MNRAS.521.1674C}. A special strategy was applied in the Gaia RVS processing to measure the radial velocity for such warm stars \citep{2022arXiv220605486B} due to the presence of Paschen lines, which, in our case, do not exist \citep[see Fig.6]{2022A&A...667A..83T}. However, from our mid-resolution 2.3m/WiFeS spectrum, we detected that some emissions were filling in the Ca II triplet absorption lines. Such weak emission was also seen with H$_{\alpha}$. We strongly suspect that the large RVamp value is incorrect and most likely due to some inaccurate measurements, most certainly resulting from the split of the three Ca II lines indirectly produced by the emission. Moreover, the error on the median RV value reported is about 14 km s$^{-1}$, which is not coherent with the scale implied by the RVamp.

The number of visibility periods for the remaining 68 bright stars mostly ranges from two to 17, with a peak at eight. Among the HdC stars, RY Sgr still has the largest recorded RV amplitude, 24 km s$^{-1}$, followed by ASAS-RCB-6 at 20 km s$^{-1}$. Similarly to RY Sgr, ASAS-RCB-6 shows large photometric pulsations in the visible, with a total amplitude that can reach up to $\sim$0.6 mag at maximum brightness. The largest RVamp value (except for that of F152) belongs to an EHe star, V354 Nor, with 29 km s$^{-1}$. We note, though, that this star was also reported with a median RV error ( $\sim$33 km s$^{-1}$) that is higher than its RVamp value. This is likely explained by differences in the measurements used in the time series to calculate both parameters. Some obvious outliers have been removed to compute RVamp. However, the resulting RVamp value still needs to be used carefully because V354 Nor has previously been observed twice, and both results indicated a lower velocity amplitude. From the first observation, \citet{1987MNRAS.226..317J} reported eight measurements over a span of 350 days, but no significant RV variability was found, despite their capability to detect velocity variations down to 20 km s$^{-1}$. From the second observation, \citet{1993MNRAS.265..351L} measured a velocity amplitude of the order of 10 km s$^{-1}$ with nine observations made over 55 days. Gaia DR3 has observed V354 Nor on 14 transits over eight visibility periods covering 770 days. Three other HdC stars, namely B564, NSV 11154, and ASAS-RCB-16, were found with an RVamp value lower than the error on RV. Therefore, their respective results should be treated with extra caution.

We have checked the consistency of the RVamp values in Gaia DR3 against those found in the literature, particularly those listed by \citet{1997MNRAS.285..266L}. There is a clear agreement with the high amplitude recorded for RY Sgr, as well as for most of the other RCB stars whose RVamp values are typically in the range of 10-15 km s$^{-1}$. Additionally, the low RV amplitude of RT Nor reported by \citet{1997MNRAS.285..266L} still holds. Gaia DR3 reported a low value of $\sim$8.6 km s$^{-1}$ after six visibility periods. However, we have also observed some discrepancies that can be explained by the number of visibility periods used in both studies. For example, \citet{1994MNRAS.271..919L,1997MNRAS.285..266L} found a small RV amplitude of around 5 km s$^{-1}$ for UX Ant based on seven measurements taken over two months, whereas Gaia DR3 found a total amplitude of approximately 15 km s$^{-1}$ by doubling the number of measurements over a much longer period of time. We also observed the inverse with GU Sgr and V CrA, which respectively had only five and four visibility periods in Gaia DR3. With a low number of observations, RVamp will (almost) always be an under-estimate of the real value.


\section{Evolution of radio velocity amplitude and dust production rate with effective temperature \label{sec_RVamp_dust_temperature}}

A new classification system for HdC stars was defined by \citet{2023MNRAS.521.1674C} based on the spectroscopic characteristic of the stars. The system sorts the stars into eight temperature classes ranging from zero to seven, with the higher values corresponding to colder HdC stars. We used this class index as a proxy for each HdC effective temperature. For simplicity in our study, we assigned an index value of -1 for all EHe stars, as they are of warmer temperature.

\subsection{Radio velocity amplitude and effective temperature}

\begin{figure*}
\centering
\includegraphics[scale=0.50]{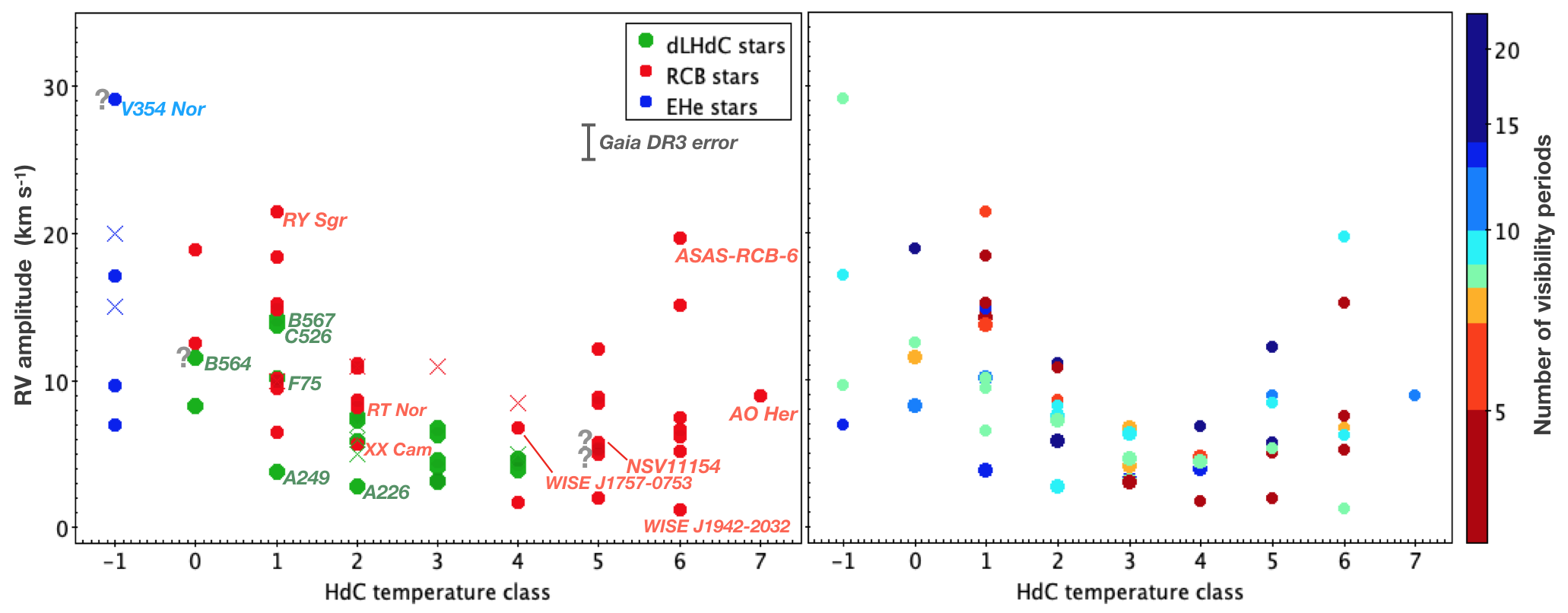}
\caption{Radial velocity peak-to-peak amplitude versus the temperature class of HdC stars determined at maximum brightness. EHe stars have been given a class of -1. In the left panel, the colour code indicates the type of supergiant star. The crosses are measurements obtained from the literature (see text), while the large dots are from Gaia DR3. The names of interesting targets with low or high RV amplitude have been added. The names WISE J1942-2032 and WISE J1757-0753 are abbreviations for the RCB stars WISE J194218.38-203247.5 and WISE J175749.76-075314.9. The typical Gaia DR3 error on RV amplitude for bright stars is shown near the colour legend. Stars presenting an RVamp value lower than their respective error on RV are indicated with a question mark (see discussion in Sect.~\ref{subsec_IntrinsicRVvar}). The right panel shows the same graph but colour-coded for the number of visibility periods for the stars observed by Gaia DR3. The stars with a low number are in dark red, while the ones with more than ten visibility periods are in blue.}
\label{fig_RVamp}
\end{figure*}

In Figure~\ref{fig_RVamp} (left), we plot RVamp versus the HdC temperature class. In the figure, we use interrogation symbols to indicate the four stars mentioned in Sect.~\ref{subsec_IntrinsicRVvar} whose RVamp value is lower than the error on RV. In the case of the few stars for which no amplitude values were reported in Gaia, we used the ones found in the literature. For GU Sgr and V CrA, we used the amplitude values reported by \citet{1997MNRAS.285..266L}. We did not have a temperature class for V CrA, but as its effective temperature was estimated to be $\sim$6250 K by \citet{2000A&A...353..287A}, we temporarily assigned it an HdC2 class for the purposes of this analysis. We also include the results reported by \citet{1997MNRAS.285..266L} for the four bright dLHdC HD stars  HD 137613, HD 148839, HD 173409, and HD 182040 (HdC temperature index of 4, 2, 2, and 2 resp.) with RVamp$\sim$5-6 km s$^{-1}$ (which is an upper limit), as well as the results for the RCB star HD 175893 (HdC4) with RVamp$\sim$8 km s$^{-1}$. Finally, R CrB, an HdC1 star, was reported to have a peak-to-peak RV amplitude of $\sim$10 km s$^{-1}$ by \citet{2019MNRAS.482.4174F}, and \citet{1987MNRAS.226..317J} observed amplitudes of $\sim$15 and $\sim$20 km s$^{-1}$ for the EHe stars V2244 Oph and V2076 Oph. Conversely, we note that 14 RCB stars have reported RVamp values in Gaia DR3, but their temperature classifications are not available in \citet{2023MNRAS.521.1674C}. We did not include any of these stars in Fig.~\ref{fig_RVamp}, but for the four stars observed for more than eight visibility periods, we found a temperature estimate. These four stars are UW Cen, ASAS-RCB-12, OGLE-GC-RCB-1, and WISE J173737.07-072828.1, and they have the following respective RVamp values: 16.82, 7.83, 6.61, and 7.49 km s$^{-1}$. The star UW Cen is a warm RCB star with an estimated effective temperature of 7500 K by \citet{2000A&A...353..287A}, suggesting it is likely an HdC0 or 1 star. ASAS-RCB-12 is considered a mild RCB star \citep{2013A&A...551A..77T, 2023MNRAS.521.1674C}, corresponding to an HdC temperature index between 2 and 4. OGLE-GC-RCB-1 exhibits a spectrum indicative of a cold RCB star \citep{2011A&A...529A.118T}, thus endorsing indexes  equal to or higher than five. There is no available temperature information from the literature for WISE J173737.07-072828.1; however, our spectrum suggests that it is also a cold RCB star. We note that none of the other ten RCB stars exhibit notably high velocity amplitude values.

An interesting trend emerges from this graph. While there exists a large scatter in each HdC temperature class, one can observe that both the median amplitude value and the maximum amplitude recorded in each of these classes evolve with the temperature. First, we observed a decrease of the maximum from a high of $\sim$20 km s$^{-1}$ down to a level of $\sim$10 km s$^{-1}$ between the HdC6 and HdC3-4 classes followed by an increase up to 24 km s$^{-1}$ with RY Sgr, an HdC1 star. This last increase was observed with both RCB and dLHdC stars. 

To verify the validity of this trend, we examined the distribution of the number of visibility periods across the different HdC classes. Figure~\ref{fig_RVamp} (right) shows the same diagram as before but colour-coded according to the number of visibility periods. While all classes from one to six are affected by some members having a low number of visibility periods (i.e., less than five), the trend remains consistent when focussing only on stars with eight or more visibility periods. However, we noticed a possible caveat. On the colder side, only three measurements with RVamp$>$10 km s$^{-1}$ effectively make the trend we observed. Furthermore, among the mild temperature classes, the HdC3 class contains five dLHdC stars but only one RCB star (GU Sgr), so the observed trend may change as more data on RCB stars in that temperature range become available (four other RCB stars are currently known in that class). More RV measurements are therefore needed to confirm this initial observation.

The intrinsic radial velocity variation of a star is related to its internal motions, which are in turn determined by the star's internal structure and physical properties, including temperature, mass, and chemical composition. Past research \citep{1972MNRAS.158..305A,1997MNRAS.285..266L} has shown that RV variability in RCB stars seems to be driven in part by radial pulsation. However, variations in periods and amplitudes observed over time led \citet{2019MNRAS.482.4174F} to suggest that the photometric and velocity variations can be explained by the integrated effects of a small number of giant turbulent convective cells over the visible side of the star rather than a global pulsation. In the 0.7 to 0.9 solar mass double degenerate scenario, the first attempt to model irregular photometric and velocity variations in RCB stars was made by \citet{2008ASPC..391...69S}. He showed that these variations could be due to radial and long-period non-radial pulsations simultaneously excited by strange-mode instability \citep{2009CoAst.158..245S}. Only RY Sgr seems to present pure radial pulsations. According to his model, the period of these pulsations increases with decreasing effective temperature. Such a period-temperature relationship was already reported by \citet{1990MNRAS.247...91L} with most of the HdC stars they monitored, again supporting the case of radial pulsations.

We observe a phenomenon among HdC stars during their evolution from cold to hot phases. The large maximum RV amplitude values observed in cool (4000 $<$ T$_{eff} <$ 5000 K) and warm (T$_{eff} >$6000 K) HdC stars suggest the presence of favourable conditions to create broad atmospheric movement. These conditions appear to be gradually less satisfied between these two temperature ranges. There is also a significant scatter of RV amplitude values in every temperature class. This scatter may be attributed to differences in chemical compositions and masses or may simply be due to angular changes between the observer’s line of sight and the direction of the atmospheric motions. A dedicated study investigating the amplitude of photometric variations at maximum brightness would be beneficial to confirming the observed trend with effective temperature.

\subsection{Dust formation rate evolution with effective temperature of R Coronae Borealis stars}

The relationship between photometric variation due to atmospheric movement and the total amplitude of RV variations in RCB and dLHdC stars was discussed by \citet{1997MNRAS.285..266L}. The authors observed a threshold of 8-10 km s$^{-1}$ in velocity amplitude that separates these two groups of stars. They hypothesised that to create the ideal thermodynamic conditions necessary for dust formation, shock waves must be created at velocities higher than $\sim$10 km s$^{-1}$ in the photosphere. Such strong shock waves are essential in the physical mechanism presented by \citet{1996A&A...313..217W} to temporarily provide the appropriate thermodynamical conditions for dust formation close to the photosphere of RCB stars (i.e., between 1.5 and 3 stellar radii, or R$_\star$). After the shock, temperatures of the circumstellar envelope gas that is in a particular density interval can drop below 1500 K, which is appropriate for carbon nucleation, following a fast relaxation towards radiative equilibrium. They mentioned that in order to reach such gas temperatures, the shock wave velocity in the circumstellar envelope gas must respectively be at least 20 and 50 km s$^{-1}$ for radial distances of 3 and 1.5 R$_\star$. Furthermore, they underlined that small amplitude waves in the photosphere may steepen up to strong shock waves in the circumstellar envelope if the photospheric density gradient is large \citep{1996ASPC...96...83W}.
 
We effectively confirm part of the findings of \citet{1997MNRAS.285..266L}: RCB stars generally exhibit larger RV amplitudes than dLHdC stars (Fig.~\ref{fig_RVamp}), and most dLHdC stars have RVamp$<$10 km s$^{-1}$. However, there are exceptions. Two of them, F75 and C526, present higher RVamp values but were also recently seen to create some dust at a low production rate \citep[See Fig.14, 15, 17]{2022A&A...667A..83T}. On the other hand, two other dLHdC stars, B564 and B567, also exhibit such high RVamp values but no dust has been detected around them so far. We will continue to monitor their optical and IR photometric behaviour in order to detect any future dust creation events. Conversely, some RCB stars have low RVamp values (around 6 km s$^{-1}$) despite having been observed with more than nine visibility periods. Namely, these are NSV 11154, XX Cam, WISE J175749.76-075314.9, and WISE 194218.38-203247.5. They are all currently producing dust. An IR excess has been observed around all of them, and decline events have been observed in the past 10 years for all but one, XX Cam. Confirmation of the RVamp values of these targets will require additional measurements from the upcoming Gaia data release. 

\begin{figure}
\centering
\includegraphics[scale=0.46]{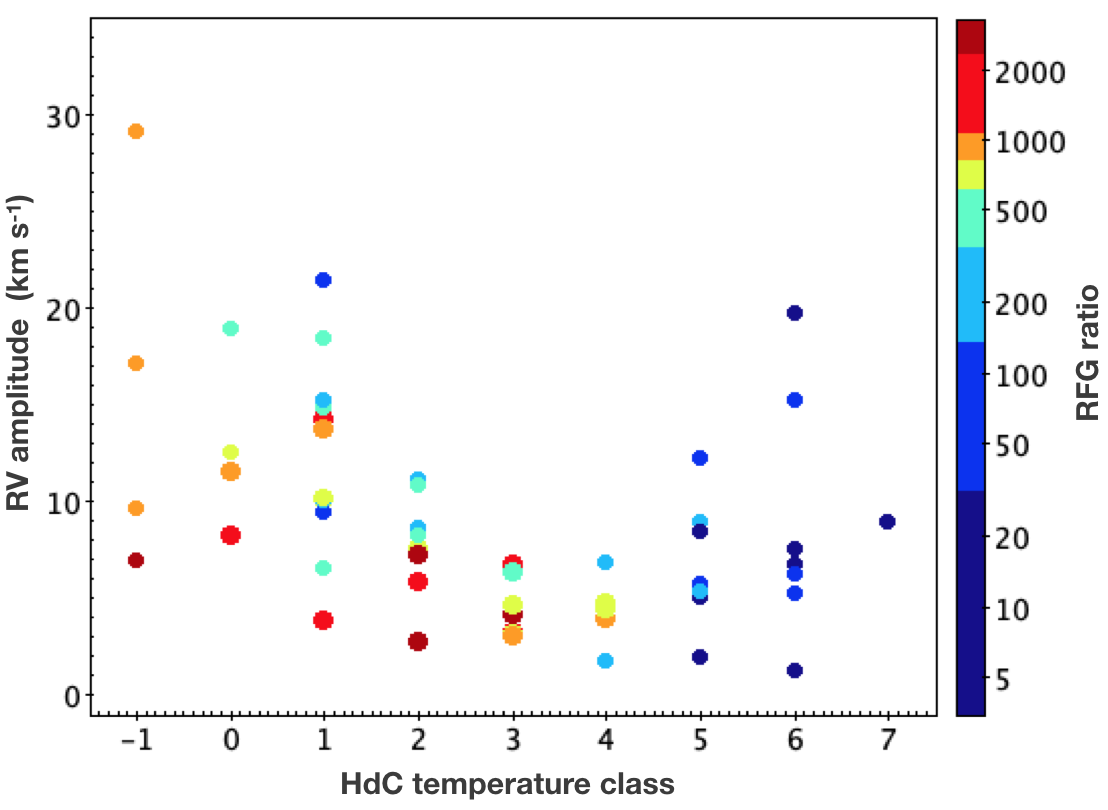}
\caption{Radial velocity peak-to-peak amplitude versus the temperature class of HdC and EHe stars determined at maximum brightness. The stars are colour-coded by the ratio RFG. A smaller value of RFG corresponds to a more active photometric decline phenomena in RCB stars. Only the bright HdC stars (GRVSmag$<$12 mag) were used.}
\label{fig_RVamp_cc-RFG}
\end{figure}

\begin{figure}
\centering
\includegraphics[scale=0.40]{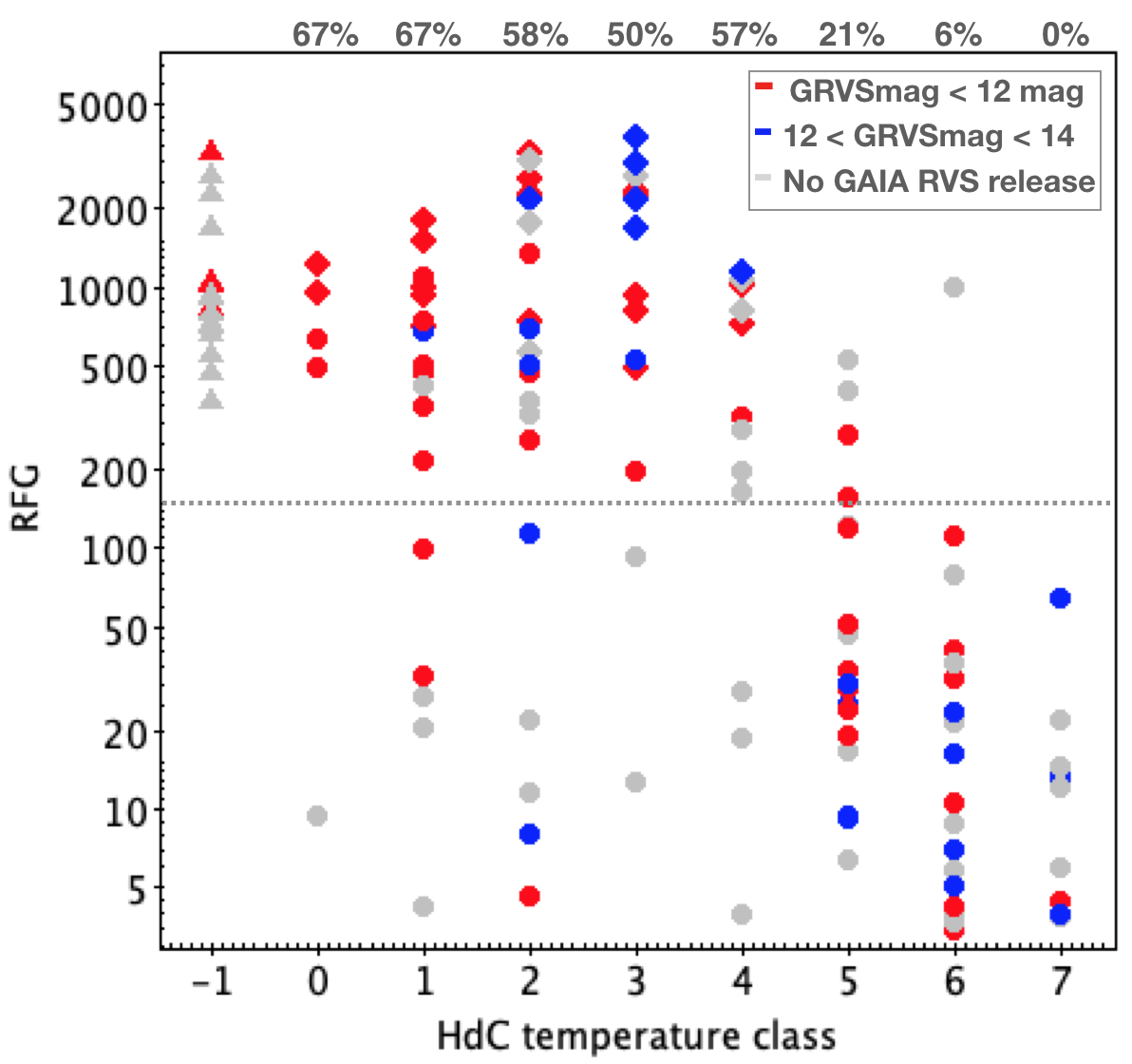}
\caption{RFG ratio versus the temperature class of HdC stars for all HdC stars whose class has been defined \citep{2023MNRAS.521.1674C}. A -1 index has been given to EHe stars. The EHe (triangles), the dLHdC (diamonds), and the RCB (large dots) stars are plotted with a colour code that shows their GRVS magnitude group. The horizontal dotted line represents the threshold (RFG$<$150) used to select RCB stars that have undergone some photometric decline phases. The percentage of RCB stars that are above this threshold are indicated at the top of the diagram for each temperature class.}
\label{fig_RFG_vs_TempClass}
\end{figure}

Despite the different RVamp values observed between RCB and dLHdC stars, there is not a well defined threshold upon which dust would be produced. Subsequently, we investigated the dependence of the dust production rate on effective temperature by using the Gaia ratio of mean G flux to its standard error (RFG) as a proxy, which is a good indicator for revealing RCB stars that have undergone decline events during Gaia DR3 (see \citet{Tiss2023b} for a detailed discussion). A smaller value of RFG corresponds to more active photometric decline phenomena. In Figure~\ref{fig_RVamp_cc-RFG}, we show a diagram similar to Figure~\ref{fig_RVamp} but colour-coded with the RFG ratio. Interestingly, we observed that the cold RCB stars (HdC 5 to 7) experience decline events more frequently. Here, the sample we used was limited to bright targets (GRVSmag$<$12 mag), as Gaia RV variability indices were only available for those stars. However, we found that our finding is also confirmed when including fainter RCB stars and all other RCB stars that did not have any Gaia RVS parameters released in our study. Figure~\ref{fig_RFG_vs_TempClass} shows the RFG ratio versus the HdC temperature class for all HdC stars. The proportion of HdC stars with a low RFG ratio is higher in the cold HdC classes than in the warmer ones: respectively, 90\% and 40\% of the cold and warm HdC stars are below the RFG threshold of 150 used in \citet{Tiss2023b} to select RCB stars that underwent a photometric decline during Gaia DR3. 

\begin{figure}
\centering
\includegraphics[scale=0.52]{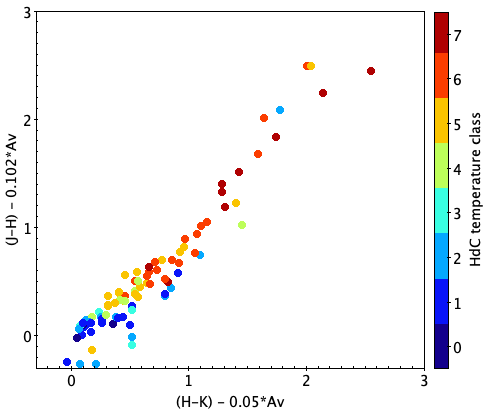}
\caption{Diagram of J-H vs H-K, after correction for interstellar extinction, for all Galactic RCB stars. The stars are colour-coded by HdC temperature class.}
\label{fig_JHvsHK_ccTempclass}
\end{figure}

Furthermore, we used the 2MASS dataset \citep{2006AJ....131.1163S} to test this result. The 2MASS survey has observed all RCB stars at different epochs, which could be considered as random for each star individually. We plotted the J-H versus H-K diagram colour-coded by the HdC class temperature (Figure~\ref{fig_JHvsHK_ccTempclass}). In the figure, the RCB stars are distributed along a curved line corresponding to all combinations of two blackbodies in various proportions, from all "star" to all "shell" (i.e., when the stars are fully obscured \citep{1997MNRAS.285..339F}). The high J-K colour index values correspond to elevated levels of infrared excess, which result from the presence of a substantial amount of dust along the line of sight of RCB stars. The RCB stars observed at epochs of maximum brightness are situated in the bottom-left side of the diagram, while those observed during the depth of their declines are located in the top-right side. We observed a clear correlation between the state of obscuration and the RCB star temperature. The RCB stars observed with a high level of infrared excess are more likely to be the coldest ones. This supports our initial observation that the cold RCB stars are more likely to be in a decline phase at any given time than the warmer RCB stars.

We have observed indirect signs of a higher rate of dust formation events in cool HdC stars, it is therefore reasonable to expect that very cold RCB stars highly enshrouded by dust could exist and remain undetected. An example of such an RCB star could be EROS2-SMC-RCB-4, which has not yet been observed in a bright phase \citep{2009A&A...501..985T,2020A&A...635A..14T}.

Our observations suggest that the rate of dust production decreases as the effective temperature of HdC stars increases. This would indicate that the thermodynamic conditions necessary for producing dust particles become less favourable as the temperature increases and the atmosphere shrinks in size. Nevertheless, dust production activity is still possible in RCB stars with warm atmospheres, as they are seen to undergo decline events, although the occurrence of these events appears to be less frequent.

\section{Conclusion \label{sec_concl}}

Gaia DR3 has provided valuable insights into the velocity variations of HdC star atmospheres. Our findings confirm the initial observations made by \citet{1997MNRAS.285..266L} that RCB stars exhibit larger total RV amplitudes compared to dLHdC stars. However, there is no clear threshold separating these two subgroups of HdC stars. 

We identified an interesting trend in the maximum RV amplitude observed within each HdC temperature class. The amplitude initially decreases from HdC6 to HdC3 and 4, but then rises again with warmer temperatures. This trend requires further confirmation from upcoming Gaia data releases as well through an increased number of visibility periods. If confirmed, this trend would suggest the existence of a general phenomenon during the evolution of HdC stars from a cold to a warm phase. The helium shell-burning giant stage seems to start with strong atmospheric motions that gradually diminish in intensity until reaching an effective temperature between 5000 and 6000 K. Subsequently, the atmospheric motions regain strength as the HdC star's atmosphere shrinks at warmer temperatures.

In relation to the dust production rate, we also observed a correlation with the effective temperature. Using two proxies to assess the photometric status of RCB stars, namely,  one from Gaia and another from the 2MASS survey, we noted that cooler RCB stars are more likely to be in a decline phase at any given time compared to warmer RCB stars. Consequently, the dust production rate appears to be highest during the cold stages of HdC stars, particularly within the HdC 5 to 7 temperature classes. In contrast, at warmer temperatures, the thermodynamic conditions required for producing dust particles seem to occur less frequently.

It is challenging to reconcile our observations regarding the trend of RV amplitude and the dust production rate with effective temperature. In the study of \citet{1997MNRAS.285..266L}, the authors proposed a necessary threshold of around 10 km s$^{-1}$ in RV amplitude in the photosphere to initiate shock waves that would generate conditions suitable for dust formation, while \citet{1996A&A...313..217W} presented a physical mechanism in which shock waves in the circumstellar envelope of even higher velocities were necessary. The new series of Gaia RV measurements suggest that this phenomenon is more complex. Additionally, the situation is further complicated by the observed differences in chemical composition \citep{2000A&A...353..287A, 2011MNRAS.414.3599J,2022A&A...667A..85C} and luminosity \citep{2022A&A...667A..83T} among HdC stars, which can influence internal motions. Therefore, a more detailed analysis of the physical properties of these stars is necessary to elucidate the underlying mechanisms driving their RV variability and, consequently, their dust production rate.

\begin{acknowledgements}

We thank Fr\'ed\'eric Arenou for his insight on the Gaia datasets. PT personally thanks Tony Martin-Jones for his usual highly careful reading and comments. PT acknowledges also financial support from “Programme National de Physique Stellaire” (PNPS) of CNRS/INSU, France. AJR was supported by the Australian Research Council through award number FT170100243. We also thank the team located at Siding Spring Obser- vatory that keeps the 2.3m telescope and its instruments in good shape, as well as the engineer, computer and technician teams located at Mount Stromlo Observatory that have facilitated the observations.

This work has made use of data from the European Space Agency (ESA) mission Gaia (https://www. cosmos.esa.int/gaia), processed by the Gaia Data Processing and Anal- ysis Consortium (DPAC, https://www.cosmos.esa.int/web/gaia/dpac/ consortium). Funding for the DPAC has been provided by national institutions, in particular the institutions participating in the Gaia Multilateral Agreement. This publication makes use of data products from the Two Micron All Sky Survey, which is a joint project of the University of Massachusetts and the Infrared Processing and Analysis Center/California Institute of Technology, funded by the National Aeronautics and Space Administration and the National Science Foundation. This research has made use of the SIMBAD database,operated at CDS, Strasbourg, France.

\end{acknowledgements}

\bibliographystyle{aa}
\bibliography{GAIA_RV_Variation}


\end{document}